# ECONOMIC CONSEQUENCES OF COVID-19 PANDEMIC TO THE SUB-SAHARAN AFRICA: AN HISTORICAL PERSPECTIVE

**Akinlo, Anthony Enisan**
Department of Economics
Obafemi Awolowo University, Ile-Ife
Email: aakinlo@oauife.edu.ng

**Ojo, Segun Michael**
Department of Economics
Redeemers University, Ede
Email: ojosegunm@yahoo.com

**ABSTRACT**

This paper examines the economic consequences of the COVID-19 pandemic to sub-Saharan Africa (SSA) using the historical approach by analyzing the policy responses of the region to past crises and their economic consequences. The study employs the manufacturing-value-added share of GDP as a performance indicator. The analysis shows that wrong policy intervention to past crises, lead the African sub-region into the deplorable economic situation. The study observed that the region leapfrogged prematurely to import substitution, export promotion, and global value chains. Based on these past experiences, the region should adopt a gradual approach in responding to the COVID-19 economic consequences. The sub-region should first address relevant areas of sustainability, including proactive investment in research and development to develop home-grown technology, upgrade essential infrastructural facilities, develop security infrastructures, and strengthen the financial sector.

**INTRODUCTION**

The outbreak of COVID-19[1] pandemic is relatively unique both in presentation and magnitude (Boissay and Rungcharoenkitkul, 2020). Therefore, the economic consequences of the COVID-19 pandemic are incomparable to that of the two major past disease-related crises namely; the human immunodeficiency virus (HIV) and Acquired immunodeficiency syndrome (AIDS) popularly known as HIV/AIDS, and the Ebola outbreak of 1976 (CDC, 2016). According to International Monetary Funds (IMF) (2020), the COVID-19 pandemic is the worst global crisis since the Great Depression of the 1930s. HIV/AIDS and Ebola outbreaks did not result in a global-crisis, lockdowns, and paralyzing of economic activities in virtually all countries simultaneously like the case of the COVID-19 pandemic. So, the crises that have the similitude of COVID-19 were the First World War, The Great Depression of the 1930s, and the Second World War. Although these are not disease-related problems, they have the same magnitude of impacts as COVID 19. COVID-19 is not a sectional or regional crisis, it is a pandemic of global magnitude (Gennaro, Pizzol, Marotta, Antunes, Racalbuto, Veronese, and Smith, 2020). The supply and demand-side shocks that accompanied the COVID-19 pandemic, as a result of total and sudden paralyzing of the economy are unprecedented in the world's recent economic history. Therefore, it may likely bring about a change in the world economic order as it happened in the wake of the First World War, The Great Depression, and the Second World War.

The economic disruptions that accompany the COVID-19 pandemic involve the diversion of countries' national expenditures from other sectors to health, food, and security. In other words, the COVID-19 pandemic brought about the sudden divert of the government's spending to medical and paramedical sectors, food supply, and security infrastructures due to lockdown operations. Therefore, one positive thing about the COVID-19 pandemic is that it brings about the upgrading of the medical facilities, reinforcement of food supply capacity, and strengthening of security operatives in many nations. These will exert some positive impacts on the economies of the countries. But the havocs the pandemic has wrecked on nations in terms of the loss of lives, unplanned medical and burial expenditures, unemployment burden, business sector's revenue loss, and the general disruptions of economic activities; outweigh its positive features. Consequently, the impending global economic recession may be inevitable.

---

[1] Coronavirus disease 2019 (COVID-19) is referred to as an infectious disease caused by a novel coronavirus also known as severe acute respiratory syndrome coronavirus 2 (SARS-CoV-2; formerly called 2019-nCoV). It was first noticed in connection with an outbreak of respiratory illness cases in Wuhan City, Hubei Province, China.

**HISTORICAL BRIEFING ON THE WORLD ECONOMY AND GLOBAL CRISES (MID-18TH CENTURY TILL DATE)**

The well-organized present world economic system did not start in the manner in which it operates today. The world economy has gone through a series of evolutionary developments that cumulate to the modern economic system. In other words, the modern economic system as well as the present world economic order are products of historical events. The successes and failures that trailed the responses of nations and groups of nations to the aftereffects of some past notable crises brought the world to where it is today. Therefore, SSA can learn from history regarding how to carry out proactive post-COVID-19 economic recovery plans. The failures of the SSA countries to respond appropriately to crises have severally thrown the region into stagnation and perpetual low economic performances. So, it is high time the region remembered history to prevent it from being repeated.

The modern international economic order took its root from the economic revolution of the mid-18th century. Before this period, the majority of the nations' economies were characterized by subsistence living. In other words, the world economy was significantly enhanced by the breakthrough in communication and transportation infrastructures of the 18th century which boosted the exchange of goods and services among the nations of the world (Shafaeddin, 1998). The economic bonds that ensued from the resultant economic interdependence significantly enhanced the efficient utilization of the world spatially distributed economic and natural resources. The development helped to instill the widest practice of comparative advantage paradigm in the international economic scene for the first time. The further scientific and technological discoveries that came up in the late 18th century and early 19th century laid the foundation for Britain's industrial revolution of the 19th century, which pioneered and motivated industrialization in other industrialized nations.

Before the outbreak of the First World War, the world economy was united by the operation of the gold standard and the attendant international trade. The first global crisis that truncated the first world economic order under the gold standard was the First World War. The war broke out on a world economic system that was relatively stable with smooth flows of transactions across the nation's boundaries. Put differently, the gold standard worked well till the outbreak of the First World War in 1914. However, the gold standard mechanism could not control the economic complications that arose in the international economic scene as a result of the war. The severity of the war on the global economy was consequential to the extent that it put the existing world economic order in disarray. Consequently, the gold standard mechanism lost its grip and it was abandoned for a free exchange rate system whereby individual countries reversed back to their national currencies. The war revealed among other things that, the cooperation and mutual understanding shared by the trading countries was the force behind the gold standard mechanism. After the war, the free exchange rate system that took the place of the gold standard failed to restore the pre-war world economic order. Because the free exchange rate system was characterized by strong competition among the countries and their currencies. However, Britain re-embraced the gold standard and all the colonies under the British Empire followed suit. The British group has a large membership, and that helped them to prevail over other sub-groups to secure global acceptance of the gold standard for the second time. Afterward, Britain gained economic dominance over other countries and so the British pound sterling. Therefore, pound sterling and gold were the dominant currencies, and countries around the world were holding foreign reserves as gold and pound sterling (Dwivedi, 2009).

The new gold system could not restore the stable pre-war international economic system, due to speculations and precautionary measures taken by the countries. However, the British economy ran into a crisis. And the development led to the British loss of her gold to other countries like France and other European countries.

Consequently, the British government withdrew from the gold standard because she had lost her take in gold. As usual, the Commonwealth of Nations imitated Britain by jettisoning the gold standard which led to total abandonment of the gold standard in the world economic system. In 1929 the U.S stock market crashed. The incident led the world into the Great Depression of the 1930s due to inappropriate policy responses. The world was barely recovering from the 1930s Great Depression when the Second World War broke out due to some political complications. In the early 20th century, the world economy was traumatized by three global calamities namely; the First World War, the 1930s Great Depression, and the Second World War.

The post-World War II era was another action-parked phase in the world economic history. The period witnessed two geopolitical phenomena; the Cold War and the decolonization of former colonies. The economic confusions that trailed the war also motivated the inauguration of notable international trade and financial institutions- the International Bank for Reconstruction and Development (IBRD) or the World Bank, International Monetary Fund (IMF) and international trade organization (ITO) which later metamorphosed to World Trade Organization (WTO). Those institutions were initiated by the world economic managers to put a new economic order in place to prevent the reoccurrence of the events that led to the Second World War and the Great Depression. Each of the international institutions has primary responsibility for which it was created. The International Bank for Reconstruction and Development (IBRD) or the World Bank was established to rebuild the war-battered economies particularly in Europe and Asia, the two regions that bore the brunt of the war. International Monetary Funds (IMF) was saddled with the responsibility of maintaining a fixed exchange rate system called the Bretton Woods System. The Bretton Woods exchange rate system collapsed in the early 1970s. The International Trade Organization (ITO) initially did not materialize due to some political factors but later restructured as World Trade Organization (WTO) with the sole assignment of liberalizing international trade at the global level.

The developed countries that were battered by the war had a rapid recovery, such that their economies stabilized within the first decade after the war while the sub-Saharan African economies were stagnant. The concern over the developmental challenges of the developing countries drew the attention of policymakers and the development partners first in the post Second World War. Before this time, the majority of the developing countries were colonies of the developed countries who used them as sources of raw materials and cheap labour for industrial activities in their home countries. However, it was common knowledge that all the developed countries at their takeoff did rigorous industrialization and manufacturing sector development. Consequently, industrialization was recommended as a catch-up strategy for developing countries. The advocacy for industrialization and manufacturing output development as a catch-up strategy for the developing countries (Prebich, 1950; Singer, 1950), was reinforced by the groundbreaking publications by Nicholas Kaldor in the mid-1960s known as Kaldor's law. The Kaldor's law is in two forms namely; the Kaldor's first and the Kaldor's second law (or the Verdoorn's law) (Kaldor, 1966). The first law says that manufacturing is the engine of growth, while the second law postulated that there is a positive causal relationship between output and labour productivity in manufacturing derived from static and dynamic returns to scale.

The wide acceptance of this theory reinforced the adoption of industrialization and manufacturing output development as a catch-up strategy for developing countries through import substitution policy (Lin, 2011). Besides, the pursuit of import substitution was partly triggered by the maltreatments the developing countries went through in the hands of their colonists. The primary products of the colonies were exported to the colonists' countries for industrial uses, and consequently, the colonists did not promote industrial development nor infrastructural upgrading in their colonies. Thus, when the colonies were decolonized; the consensus was that they should stop the exportation of their raw materials. This implies that the developing countries (colonies) would utilize their raw materials locally and develop their industrial sectors thereby. Hence, import substitution was adopted in a bid to prevent neo-colonization and further resource exploitation.

Throughout the 1960s and 1970s, import substitution was the industrialization policy tool that held sway till the early 1980s when the import substitution policy was replaced with export promotion strategy following the failure of import substitution measures (Bruton 1998, Bennett, Anyanwu and Alexanda, 2015). The import substitution agenda failed due to low value-added (the import substitution industries were putting final touches to imported goods and materials), coupled with the market limitation to low-technology commodity production where the African sub-regions import substitution agenda made little progress (IBRD, 1971). Usually, import substitution is constrained by the size of the domestic market except if the products are exportable. Unfortunately, the import substitution products in sub-Saharan Africa were not internationally competitive, hence the collapse of import substitution scheme in the region (Krueger, 1974; Krugman, 1993; Soludo, Ogbu and Chang 2004).

However, another school of thought emerged in the late 1970s that advocated for foreign direct investment (FDI) as the appropriate means of spurring industrialization and manufacturing sector development through technological transfer. Koizum and Kopecky (1977) were the first to model FDI and technology transfer.

In 1982, another landmark publication that tremendously reinforced the plausibility and theoretical relevance of the work of Nicholas Kaldor was released by Gavin Kitching, who argued that "if you want to develop you must industrialize" (Kitching, 1982). Kitching based his argument for industrialization as the catch-up strategy on the limitation of agriculture and the efficiency of economies of scale; he laid the foundation for the structural transformation approach that constitutes the main policy thrust of the late industrialization campaign (UNIDO, 2015). Therefore, the work of Kitching helped to reaffirm the plausibility of the adoption of industrialization and manufacturing development as the catch-up strategy for the developing economies.

As a result, FDI expansion and export promotion targets fostered under the defunct Structural Adjustment Program (SAP) were the policy measures for industrialization till the collapse of SAP and its inbuilt policy measures in the early 1990s. The region's manufacturing sector is characterized by low value-added firms and the export industries engaged in semi-processing primary products (Ebenyi, Nwanosike, Uzoechina and Ishiwu 2017; Bennett, Anyanwu and Alexanda, 2015). Moreover, it was further argued that local firms that lack the necessary absorptive capacity for advanced technology and skills cannot benefit from the FDI knowledge spillover (Blomstrom and Kokko, 2003). Besides, the high wave of globalization and the high intensity of the cross-border trades that followed the technological breakthroughs in the areas of ICT, mobile communication technologies, and online transactions had turned the world into a global village. This has now opened the way for the global value-chains campaign which argues for the global fragmentation of production activities (Timmer Marcel, Erumban, Los, Stehrer, and Vries, 2014), starting from the early 1990s till date.

**RECAP ON SUB-SAHARAN AFRICAN INDUSTRIALIZATION DRIVE**

This paper emphasizes the industrialization performance of SSA because the industrialization agenda is central to the economic pursuit of the region. The industrialization scheme is the only policy intervention the SSA countries have unanimously carried out in response to past crises. These include the 1973 Oil Crisis Recession, 1980 Energy Crisis Recession, 1981 Iran Energy Crisis Recession, and 1990 Gulf War Recession. Besides, goal 9 of the Sustainable Development Goals (SDG) advocates for inclusive and sustainable industrialization as a means to create employment and generate income towards achieving the desired sustainable development.

The pursuit of the industrialization agenda in SSA started with the adoption of import substitution policy in the 1960s (Mendes, Bertella, and Teixeira, 2014). Import substitution is an industrialization policy that seeks to replace imports with domestic production. However, the question is; is import substitution the right policy tool for the sub-Saharan African countries at that time? These newly borne nations lack education, road, electricity, intuitional infrastructures, and other relevant facilities. Despite that, the experts advised the nations to go on the import substitution agenda, and they complied. The proponents of the import substitution agenda based their argument on the natural resources that some of the SSA countries possessed. However, less is thought about the machine equipment, the energy consumption, transport facility, road access, the technical know-how, financial capital, and the institutional arrangements required to foster a successful import substitution agenda.

Figure 1 shows the trend in the manufacturing value-added share of GDP in sub-Saharan Africa from 1965 to 1985, which was the period of import substitution scheme. Coincidentally, the two past major epidemics broke out within the period of the import substitution scheme. In other words, the earliest case of HIV was detected in 1959 from a man in Kinshasa, the Democratic Republic of the Congo. While the Ebola virus was first discovered in 1976 in the same Democratic Republic of Congo. The two diseases did not engender serious economic disruptions like COVID-19, because the two diseases did not break out drastically like COVID-19. Besides, HIV/AIDS and Ebola did not spread out throughout the period. Consequently, their economic impacts were not too noticeable. This study employs import substitution policy to explain the responses of the sub-Saharan African countries, because import substitution policy was the economic policy that unified the SSA at that period. Besides, the most accessible economic performance indicator for that period is manufacturing value-added share of GDP. Therefore, this study adopts manufacturing value-added share of GDP as performance indicator to explain the economic performances of the SSA during these periods.

In the late-1960s when most of the sub-Saharan African countries embraced import substitution policy, the countries recorded positive growth. But, the improvement in MVA growth was short-lived due to the failure of the indigenous firms to sustain growth-enhancing productivity in the infant nations. The situation was relatively better during the 1970s when the oil-exporting sub-Saharan African countries had windfall income as a result of the oil crises of the 1970s which favoured the oil-exporting developing countries. While the oil-importing developing countries were able to secure huge loans for a bailout from the savings made by their oil-exporting counterparts. Consequently, the region had little relief from the early to mid-1970s. In sum, the

import substitution policy did not know success because it was a wrong policy response to the start-up-challenges of the newly borne nations at that time.

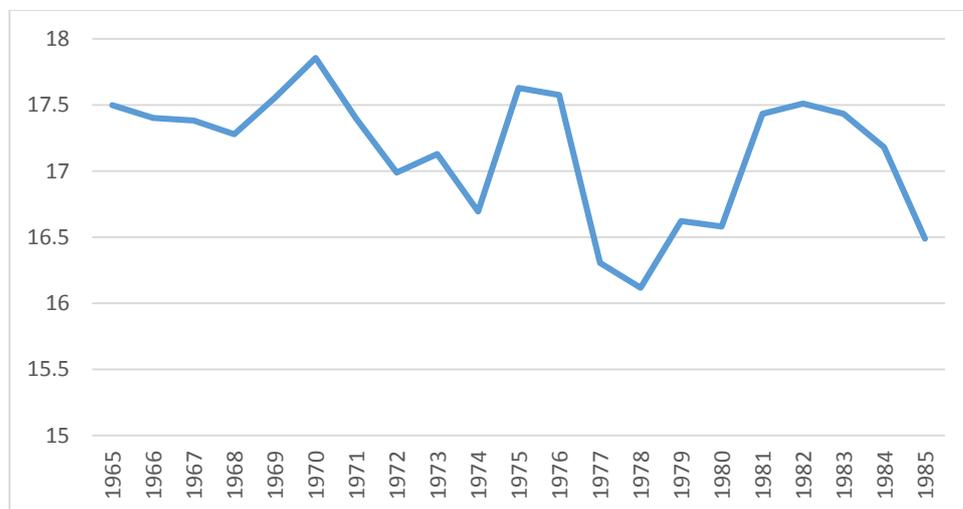

Author's computation

Figure i. **Manufacturing Value Added Share of GDP (1965-1985)**

Export-promotion-drive began in sub-Saharan Africa in the early 1980s after the two consecutive economic recessions (1980 Energy Crisis Recession and 1981 Iran Energy Crisis Recession) gave the import substitution program the last straw that breaks the camel's back (Johnson and Wilson, 1982). The period also coincided with the expansion of HIV/AIDS and Ebola epidemics. HIV/AIDS and Ebola were horrible diseases but they were not ranked as pandemic because their economic impacts were not so severe. Within the first 39 years of the HIV/AIDS outbreak 75 million people were infected and 32 million people were reportedly died from AIDS-related illness (PEPFAR, 2020). In west Africa, Ebola claimed about 11,316 lives during the 2014 epidemic. Those epidemics were not drastic like COVID-19, and sub-Saharan African countries managed them along with other economic and socioeconomic crises. The only noticeable economic consequence of Ebola occurred in Guinea, Liberia, and Sierra Leone. According to a World Bank report, about $2.2 billion was lost in 2015 in the gross domestic product (GDP) of the three countries. Germany, the U.S.A, and WHO bore the brunt of the financial burdens of the epidemic for the African countries.

As the HIV/AIDS infection started spreading in the early 1980s, the sub-Saharan Africa countries were on export promotion scheme through the 1980s structural adjustment program (SAP). The programs of action include privatization, fiscal austerity, free trade, and deregulation. The economic setback that befell the region under export promotion was more than the failure of import substitution, as evident in Figures 1 and 2. In other words, the manufacturing value-added share of GDP during the import substitution period oscillated but during the export promotion period, it plummeted. The reason for the economic retrogression under the export promotion program is not farfetched. The unsolved underperformances of import substitution policy translated to the poor outcome of the export promotion agenda. It was the success of the import substitution policy that should have facilitated the attainment of the export promotion goals and targets.

In figure 2, the manufacturing value-added curve was downward sloping from left to right throughout the export promotion period. Before the adoption of the export promotion agenda, the manufacturing value-added of the region was relatively high. The descriptive analysis in table 1 and 2 show that the minimum performance of the SSA during the import substitution period is approximately equivalent to the maximum performance under export promotion. The average performance under import substitution is far higher than the average performance under export promotion. In other words, the manufacturing productivity of SSA dissolved into a persistent decline since the inception of the export promotion scheme.

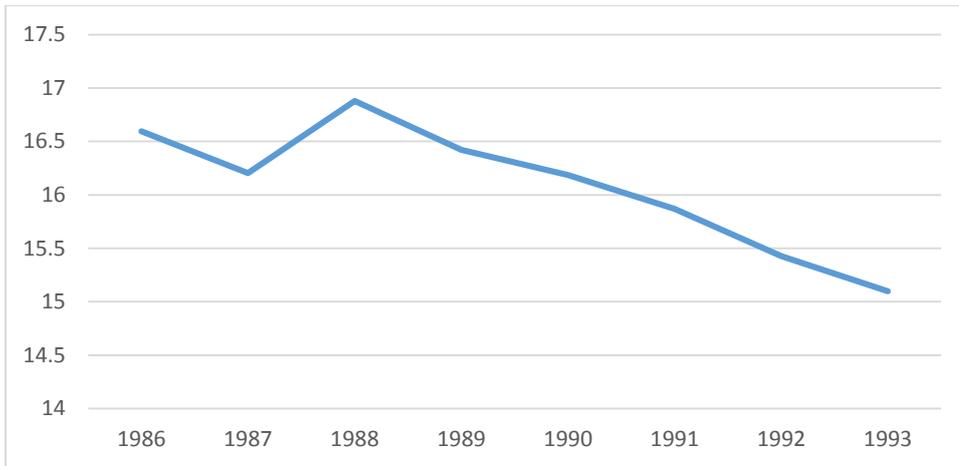

Author's computation

Figure ii: **Manufacturing Value Added share of GDP (1986-1993)**

Global value chains are the industrialization policy tool that replaced the export promotion agenda in the international manufacturing environment. The scientific and technological breakthrough in information technology in the early 1990s intensified globalization in no small measure. The resultant increase in the volume of trade and interdependence among nations stirred international fragmentation of the production process, such that countries specialize in different stages of value addition in the production process. The inability of the sub-Saharan African countries to participate competitively in global value chains, responsible for the continuous decline in manufacturing value-added since the inception of global value chains trading. It has been argued that any nation or region that fails to participate in global value chains will be marginalized in the international manufacturing markets (Moris and Fessehaie, 2014). Global value chains had held sway in the international manufacturing market since the early 1990s. The sub-Saharan African manufacturing value-added share of GDP shows that the region's level of participation in global value chains is still low. The deindustrialization trend that set in since the era of export promotion in the region does not exhibit an iota of improvement till the outbreak of COVID-19. Here is the COVID-19 pandemic; where should the region go from here? What is the right policy response for the region?

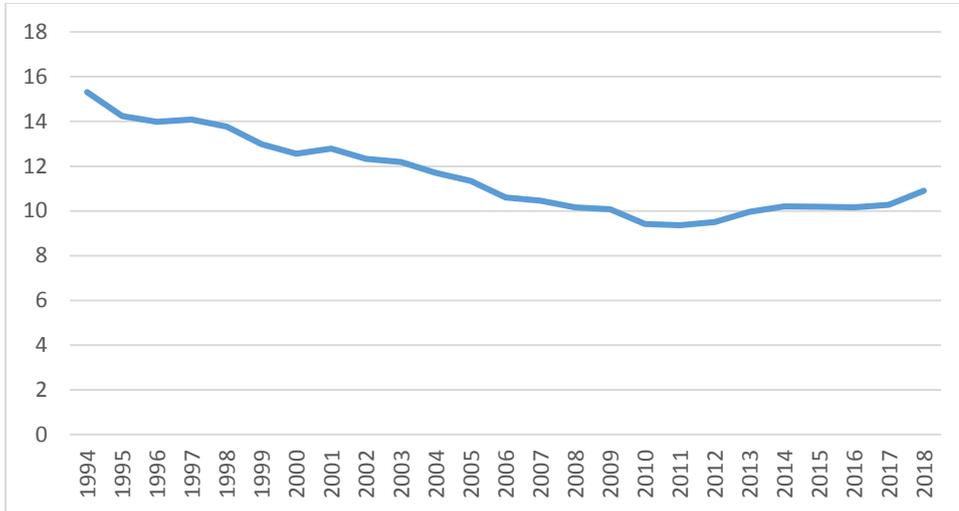

Author's computation

Figure iii: **Manufacturing Value Added share of GDP (1984-2018)**

From the foregoing historical briefing, the modern world economy has witnessed three categorical global shocks before the outbreak of the COVID-19 pandemic. The first global crisis was the First World War, followed by the Great Depression of the 1930s and the Second World War. The economic consequences of these global phenomena were aggravated and prolonged unnecessarily due to inappropriate policy responses. The uncoordinated floating exchange rate system that erupted after the First World War. The Breton woods policies and arrangements initiated in response to the economic consequences of the Great Depression and the Second World War, only made the SSA countries worse off. This is evident in the persistent economic retrogression that characterizes the region. Therefore, the main concern over the sub-Saharan African COVID-19 economic consequences; centers on the formulation of appropriate policy intervention by the relevant authorities. For instance, in the 1970s and 1980s when the sub-Saharan African countries leapfrogged to import substitution and export promotion without adequate preparation. Japan and other Asian tigers were busy building their industrial infrastructures, today they are the factories of the world.

The success of import substitution is supposed to facilitate export promotion. All the developed countries at the initial stage of their industrial development tolled the part of import substitution and protection of infant industries through restrictions. It's common to all the developed countries that after a successful and stable import substitution, they seek bilateral and multilateral trade agreements to sell their surpluses to the outside world. No developed or industrialized nation has ever embarked on export promotion without a successful import substitution (Shafaeddin, 1998). Import substitution strategy is still the best approach a country can use to nurse indigenous firms and thereby build its supply capacity that will enhance the country's competitiveness in the export promotion (Shafaeddin, 1998).

The SSA could not participate well in global value chains due to the inherent barriers in the region's economic system. Presently, the service sector is dominating the economic activities of the region. The average service value-added share of GDP in sub-Saharan Africa over the period 1981 to 2018 is estimated at 47.5% while the manufacturing value-added share of GDP is 13.1%. A country that failed on export promotion stands a narrow chance of success in global value chain trading. Consequently, the sub-Saharan African manufacturing value-added share of GDP over the past three decades is on a continuous decline due to the premature deindustrialization trend in the region. The region moved from failed import substitution to export promotion, and from unsuccessful export promotion to global value chains. When the region was supposed to be utilizing the proceeds from the exports of primary products to build the supply capacity required for sustainable industrialization, the countries leapfrogged to active industrialization through import substitution agenda. Therefore, the industrialization experiences of the sub-Saharan African countries can be described as a movement from premature industrialization to premature deindustrialization.

Another important observation from the historical review is about international cooperation failure. The world economy has witnessed several instances of international cooperation failures. The first disruption occurred in the wake of the First World War when the gold standard system that held the world economy together lost its grip due to the resultant cessation in the cooperation among the nations following the economic hardships that trailed the war. Nations could no longer listen to one another. Rather, every government was preoccupied with internal arrangements and rearrangements amidst speculative and precautionary measures in a bid to prevent the reoccurrence of the economic hardships that were inflicted on them by the unfortunate incident of the war. The immediate outcome of this scenario was the abandonment of the gold standard which has been thriving through the support of international cooperation among the nations. However, during the post-war period, there was a reemergence of the gold standard after it was abandoned for a couple of years.

The gold standard system relied on cooperation among the countries to operate a fixed exchange rate and international capital mobility. But the war put the countries apart and the economic situation across countries was worsened, individual governments introduced currency and capital controls which put end to the second era of the gold system. This marked the second abandonment of international cooperation. The situation persisted till the outbreak of the Second World War in 1939. The instability in international trade and the retrogression in the world economy caused the economic managers to initiate the Bretton Woods Agreements in 1944, to restore cooperation among the nations. Historically, whenever an incident of global impact occurred there used to be abrupt abandonment of existing international cooperation and the emergence of new world economic order. So, when the dust of the COVID-19 pandemic settles; the sub-Saharan African countries should expect a new world economic order because there would be a modification in the existing international economic cooperation if not total abandonment. Therefore, nations and regions that depend on international agreements and cooperation may end up being let down amidst the likely economic confusions that may arise in the post-COVID-19 era.

# ECONOMIC CONSEQUENCES OF COVID-19 PANDEMIC

The economic consequences of the pandemic to the SSA can be viewed from three major perspectives namely; the present and future impacts of the demand and supply shocks that result from the internal lockdown operations on the economic lives of SSA, the spillover effects of the collapse of the global economic activities on productivity and financial stability of SSA as a result of international lockdowns. The third scenario is the fall in commodity prices (particularly oil) that erupted as a result of the pandemic which is set to aggravate the economic hardship of the oil-exporting countries in the region. Figure 4 depicts the channels through which the economic consequences of the COVID-19 pandemic are transmitted in the context of the sub-Saharan African economy. On the one hand, the internal lockdowns in the countries triggered supply and demand shocks through the sudden disruptions inflicted on economic activities. The lockdown put a stay on the production of goods and services which lead to supply failure. The resultant revenue disaster engendered unplanned mass down-sizing in organizations because the employers could not keep hold of their workers in the face of zero revenue. This leads to high unemployment and partial unemployment that erupted in the course of the lockdowns. One of the immediate effects of the sudden rise in unemployment is a sharp fall in per-capital income because all the sources of generating income are choked off. The shortage in supply of goods and services together with the decrease in income caused prices of goods and services to rise and consequently dampened demand.

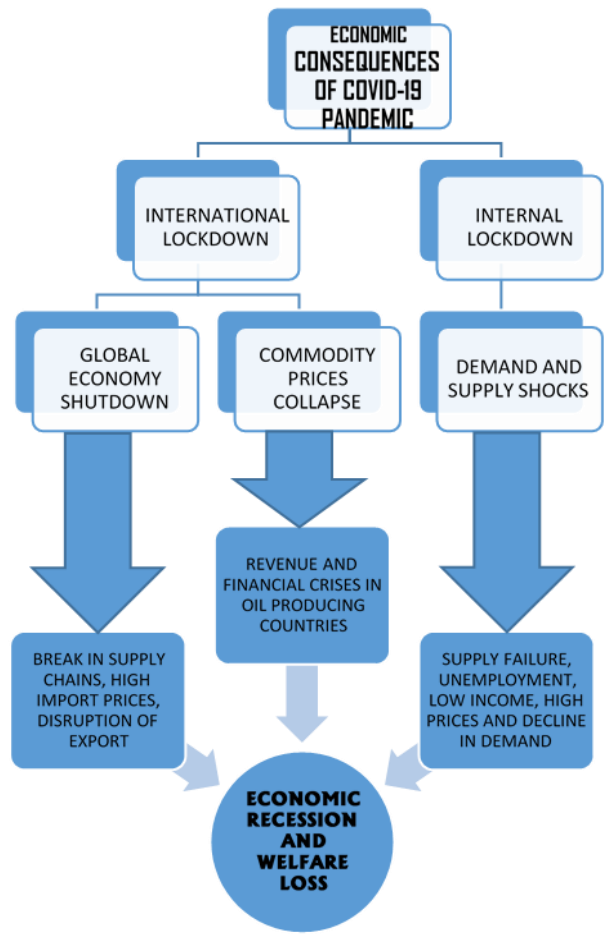

Author's computation

**Figure iv: COVID 19 Economic Consequences Impact Channels**

On the other hand, international lockdowns plummeted international economic transactions and suppressed commodity prices. The lockdown hits manufacturing industries hard and causes disruptions in global value chains and the supply of products. The break-up of the international supply chain causes imports to be dearer and aggravates the economic situations in import-led economies like SSA. Many SSA countries earn their public revenue and foreign exchange earnings from the export of one primary product or the other. The abrupt international lockdown on exports and the global oil price collapse throw many SSA countries into a serious financial disaster.

The most significant of the lessons from the COVID-19 pandemic centers on the need for an individual country's self-reliance. Whenever there is international cooperation failure, the nations' level of self-reliance becomes their parameter for survival. If a section or region is engulfed in crisis, other areas or regions that are not affected can come to the rescue of the affected areas. But in a situation whereby all countries are trapped in a calamity like the case of COVID-19, it often causes international cooperation failure which will leave every country to survive by her resourcefulness. Therefore, the pandemic is a wake-up call to the sub-Saharan African governments to realize the risk involved in unhealthy international agreements they often indulged themselves and do the needful to reposition their dysfunctional economies.

Besides, the COVID-19 pandemic helps us to identify three areas of self-reliance that is pertinent to a country's chance of sailing through a crisis like COVID-19. The areas of self-reliance include; security, food, and medical. In the economics literature, the basic needs of an individual are made up of food, shelter, and clothing. But at the national level, COVID-19 experience has revealed that the basic needs of a nation are security, food, and medical. All other sectors are lockdown and their activities suspended, but these three sectors could not be suspended. This implies that the supply capacity in those three areas is very crucial to a nation's life. The sub-Saharan African countries have a lesson to learn from this pandemic about the need to develop security infrastructure, food supply capacity, and medical facilities.

**THE WAY FORWARD**

The sub-Saharan African countries should expect a new world economic order after the COVID-19 pandemic is fully exterminated. In other words, history shows that after a global crisis like COVID-19; there is always a change in the existing world economic order. So, the COVID-19 pandemic is ongoing, SSA countries should start bracing themselves up for the policy intervention that will mitigate the negative impact of the pandemic on their economies. The region should know that the new world economic order that will trail the COVID-19 pandemic may alter the existing international cooperation. This implies that the African sub-region should countless on international support in her effort to ameliorate the negative effects of the aftermaths of the COVID-19 pandemic on her economy.

From the historical review, it was observed that SSA always seek momentary relief in the wake of the crisis. It should be clear to the region from the historical perspective that, those temporary solutions always leave their economies with prolonging economic setbacks. The focus of the governments of the region should not be merely securing IMF and world bank loans to cushion the effects of the pandemic on the less privileged in the societies. Of course, the pandemic has exerted an adverse effect on almost everyone in society, and there is a need to salvage the lives of the vulnerable members of the communities from the hit of the pandemic. But policies should pay specific attention to prevent long-term damage to the economy and also prioritize the overall economic progress of the citizens. If the temporary relief agendas like; free feeding, unemployment benefits, and health benefits are the limit of the government response to the hit of the pandemic. The region's economy will remain susceptible to future crises.

One of the facts that can be deduced from the historical briefing is that the developed and industrialized countries' recovery process is usually faster than that of the developing countries. The developed countries have structural, institutional, and social overhead buffers to absorb shocks than the developing countries who are generally weak and vulnerable. Therefore, the SSA countries should tend towards sustainability-oriented policies and programs like; investment in technological innovation and development of home-grown technologies, proactive support for the adaptation and diffusion of green technologies, industrialization and manufacturing output expansion and diversification. IMF's May 11, 2020 Report Update, advocates that policymakers should utilize the COVID-19 pandemic opportunity to carry out fundamental restructuring and repositioning of their economies. Such that future inevitable shocks will not destabilize them like the way the current one is affecting them.

The developed countries and SSA countries would need different strategies for recovery from the pandemic. Put differently, the SSA being third-world countries cannot imitate the first and the second world countries in their approaches and responses to the effects of the pandemic on their economies. The individual country has peculiar situations and it lies in each country to identify her challenges, strengths, and prospects while developing policy to remedy the present and future consequences of the pandemic on the economy. According to Shafaeddin (1998), each country having specific conditions would require specific policies at any particular period.

The SSA African should not act in a hurry in their responses to the economic effects of COVID-19. Policies and decisions that are made in haste may not yield the desired result. The governments of the SSA should patiently carry out wide consultation regarding the appropriate policy measures that will yield a lasting solution for their economies. The import substitution policy response to the post-war economic crisis in the 1960s was not appropriate. The export promotion agenda that was launched after the economic recessions of the early 1980s was also an unfortunate policy intervention. The current leapfrogging to the service-led economic system at a time when the industrialized nations are maximizing manufacturing value-added and market share in the global market equally leaves much to be desired. Therefore, the right economic decision and policy measures should be applied to the current situation in order not to allow history to repeat itself.

**CONCLUSION**

This paper preempts the likely economic aftermaths of the current pandemic in the light of historical facts and offers guidelines on how the sub-Saharan African (SSA) countries can handle their post-pandemic economic recovery plans. COVID-19 pandemic is rated as the world's worst global crisis since the Great Depression of the 1930s. However, the COVID-19 pandemic scenario is on the one hand far from being desirable, but on the other hand, the development helps to unfold the unobserved weaknesses in the SSA economic lives. The scenario answers the old saying that, the ultimate measure of a person, organization, or nation is not where she is in times of comfortability and pleasure but where she is in times of calamity. The present pandemic has revealed the true economic worth of the SSA countries. Those whose medical sectors are unfit to have realized the reality about their medical infrastructures, those who are wittingly or unwittingly over depending on foreign countries for the importation of essential goods and services have understood the height of the risk they are indulging in. COVID-19 has helped to paint the countries in their true colors. Therefore, it lies in each country to do the appropriate restructuring, transformation, reform, and repositioning were necessary to build resilience against probable future occurrences.

The paper observed that wrong policy intervention to past crises responsible in part for the economic backwardness of the region. The past policy responses to economic crises did not take proper cognizance of the weaknesses and strengths of the nations in their policy formulations and responses. In other words, the region leapfrogged prematurely in succession to import substitution, export promotion, and global value chains. This implies that in response to COVID-19 economic consequences, the region should adopt a gradual approach by first addressing relevant areas of sustainability like investing in technological innovation to develop home-grown technology, upgrade the essential infrastructural facilities, develop security infrastructures, ensure political stability, crack down effectively on corruption and all forms of illicit practices and strengthen the financial sector. This will help to provide a secure base for inclusive and sustainable industrialization and manufacturing output development that can liberate the region from over-dependence on foreign countries and set a process of sustainable development in motion in SSA.

# REFERENCES

Bennett K. O; Anyanwu U. N. and Kalu Alexanda O.U. (2015). The effect of industrial development on economic growth (an empirical evidence in Nigeria 1973-2013), *European Journal of Business and Social Sciences, Vol. 4, No. 02, May 2015. P*.

Blomstrom, M., Kokko, A., (2003)The economics of foreign direct investment incentives.*NBER Working Paper 9489.*

Boissay F. and Rungcharoenkitkul P. (2020) Macroeconomic effects of Covid-19: an early review. *Bank for International Settlements, No: 07*

Bruton, H. (1998). A reconsideration of import substitution. *Journal of Economic Literature. 36(2):903-936.*

Dwivedi D. N. (2009) Managerial economics. *NULL. ISBN: 8125923470*

Gennaro, D. Pizzol, F. Marotta, D. Antunes, C. Racalbuto, M. Veronese, V. Smith, N. (2020) Coronavirus diseases (COVID-19) current status and future perspectives: a narrative review. *Int. J. Environ. Res. Public Health, 17, 2690.*

Ebenyi G. O, Nwanosike D. U, Uzoechina B, Ishiwu V (2017). The impact of trade liberalization on manufacturing value-added in Nigeria *Saudi Journal of Business and Saudi J. Bus. Manag. Stud.*; Vol-2, Iss-5A (May, 2017):475-481

IBRD (1971), The manufacturing sector and the structure of industrial protection in Nigeria. *International Bank for Reconstruction and Development. West Africa Department*; EX-6

IMF (2020) COVID-19: An unprecedented threat to development, Sub-Saharan Africa. *Regional Economic Outlook*

Johnson, W., & Wilson, E. (1982). The "oil crises" and African economies: oil wave on a tidal flood of industrial price inflation. *Daedalus, 111*(2), 211-241. Retrieved December 15, 2020, from http://www.jstor.org/stable/20024792

Kaldor, Nicholas (1966), Causes of the slow rate of economic growth of the United Kingdom: an inaugural lecture. *Cambridge: Cambridge University Press*.

Kindleberger, C. P. (1984). Multinational excursions. *Cambridge: MIT Press*

Kitching, G. (1982). Development and underdevelopment in historic perspective. *London: Methuen.*

Koizumi, T. and Kopecky, K. J. (1977). Economic growth, capital movements and the international transfer of technical knowledge. *Journal of International Economics 7:45-65*

Krueger, A. (1974). "The political economy of rent-seeking society." *American Economic Review 64(3): 291–303.*

Krugman, P. (1993). "Protection in developing countries." In *R. Dornbusch (ed.),* (Policymaking in the Open Economy: Concepts and Case Studies in Economic Performance). *New York: Oxford University Press, 127–48.*

Lin J. Y. (2011). New structural economics: a framework for rethinking development. *The World Bank Research Observer 26(2): 193–221.*

Mendes A. P. F., Bertella M. A. and Teixeira R. F. A. P. (2014) Industrialization in sub-Saharan Africa and import substitution policy. *Brazilian Journal of Political Economy, vol. 34, no 1 (134), pp. 120-138, January-March/2014*

Morris M. and Fessehaie J. (2014). The industrialization challenge for Africa: towards commodities based industrialization path. *Elsevier Journal of African Trade 1 (2014) 25–36*

PEPFAR (2020) HIV historical timeline. *U.S. President's Emergency Plan for AIDS Relief*

Prebisch R (1950). The economic development of Latin America and its principal problems. United Nations. *New York. Reprinted in Economic Bulletin for Latin America 7(1): 1–22.*

Shafaeddin M. (1998) How did developed countries industrialize? The history of trade and industrial policy: the cases of Great Britain and the USA. *UNCTAD/OSG/DP/139*

Singer H. W (1950).The distribution of gains between investing and borrowing countries.*The American Economic Review 40(2):473–85.*

Soludo C, Ogbu O and Chang H (2004). *The Politics of Trade and Industrial Policy in Africa*. Africa World Press. Trenton.

Timmer, Marcel P., Abdul Azeez Erumban, Bart Los, Robert Stehrer, and Gaaitzen J. de Vries. (2014) .Slicing up global value chains. *Journal of Economic Perspectives,28(2), 99-118.*

UNIDO (2015). Industrial development report 2016: the role of technology and innovation in inclusive and sustainable industrial development. *United Nations Industrial Development Organization. Vienna.*